\documentclass[nofootinbib,twocolumn,showpacs,preprintnumbers,amsmath,amssymb]{revtex4-2}
\usepackage[english]{babel}
\usepackage[utf8]{inputenc}
\usepackage[usenames]{color}
\usepackage{colortbl}
\usepackage{graphicx}
\usepackage{graphics}
\usepackage{dcolumn}
\usepackage{bm}

\begin{document}

\title{Shadows and photon rings of binary black holes}
\author{S.~V. Chernov}
\email{chernov@td.lpi.ru}
\affiliation{Astro Space Center, Lebedev Physical Institute, Russian Academy of Sciences,
Profsoyuznaya str. 84/32, Moscow, 117997 Russia}

\begin{abstract}
In this paper we present the images of binary black holes using the Majumdar-Papapetrou multiblack hole solution, depending on the parameters of the problem: the mass of black holes, the distance between them, and the
inclination of the observer. The images consists of a shadows and photon rings. We find that a photon ring structure appears between black holes. The trajectories of the photons are calculated.
\end{abstract}

\maketitle

\section{Introduction}

In recent years, there has been significant progress in the physics of binary black holes. 
First of all, this is due to the discovery of the gravitational-wave background \cite{abbott2016}, which was interpreted as mergers of binary black holes of solar masses.
On the other hand, there is strong evidence for the existence of supermassive binary black holes \cite{malinovsky2023}. The main candidate for supermassive binary black holes is object OJ287 with with masses of 18 billion and 125 million solar masses respectively \cite{titarchuk2023}.
Thus, binary black holes are expected to be common astrophysical systems and the study them is of great importance in the context of general relativity.

In April 2017, the Event Horizon telescope observed shadows from the supermassive black hole in M$87^*$ and Sgr $A^*$ \cite{eht2019,eht2022}, but no one has yet observed the shadow of binary black holes.
This problem can be solved in the very near future \cite{malinovsky2023} thanks to the "Millimetron" mission \cite{kardashev2014}.
In order to do this, it is necessary to simulate the shadows of binary black holes, which is done in this work.

There are a lot of papers in which shadows of binary black holes are calculated 
\cite{nitta2011,yumoto2012,cunha2018,shipley2016}, but no one has investigated how the shadows 
depend on the parameters of the problem. In particular, how the shadow depends on the ratio of the masses of black holes, the distance between black holes, and the angle of inclination of the observer relative to black holes. There is also no work on the study of photon rings around binary black holes. In this paper, using the Majumdar-Papapetrou multiblack hole solution, shadows and
photon rings around binary black holes are constructed depending on the parameters of the problem.

In this paper, we use the geometric units, $G=c=1$.

\section{Circular orbits of binary black holes}

In this section, we will briefly describe the properties of the metric Majumdar-Papapetrou.
This metric describes a multiblack hole solution. Each such black hole has an electric charge equal to its mass. In case of a single black hole, the Majumdar-Papapetrou solution reduces to extreme Reissner-Nordstrom solution.

\begin{figure*}
\includegraphics[scale=0.5]{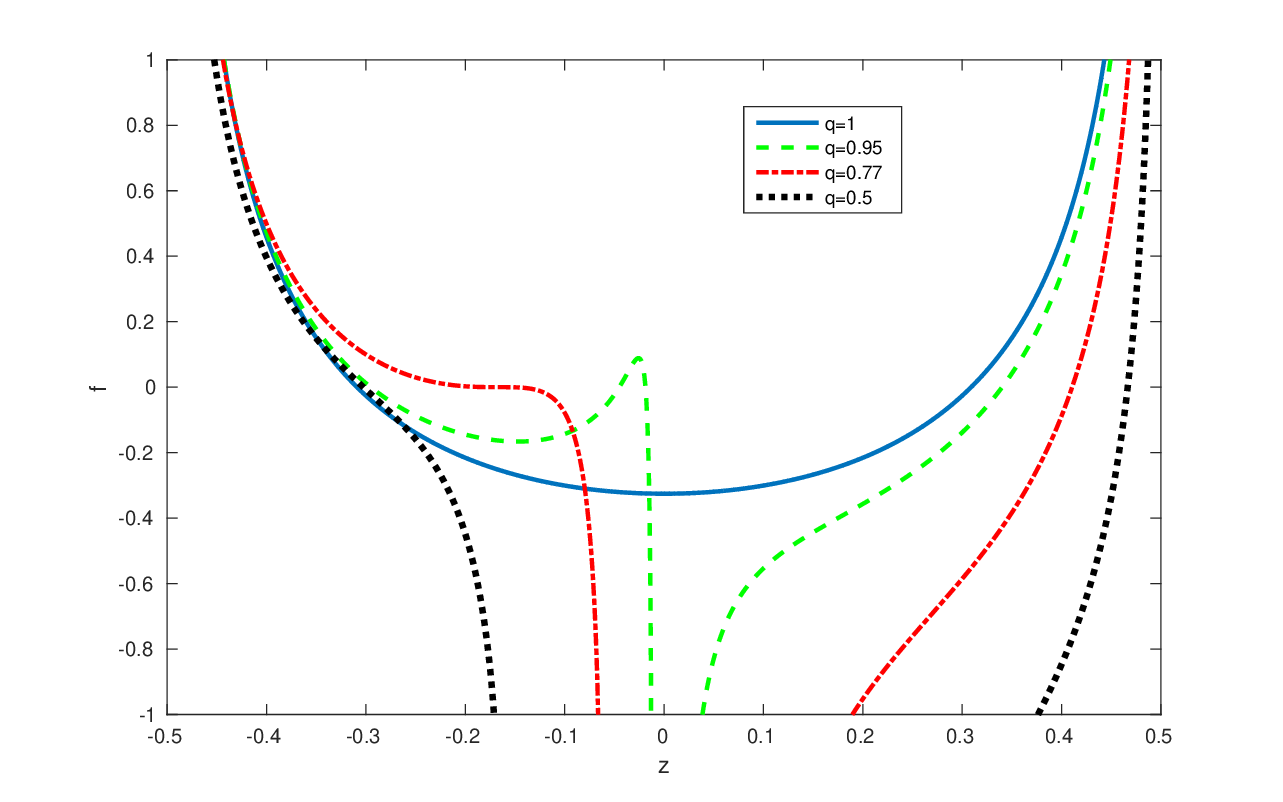}
\caption{The function f vs z for $l=0.5$ and $q=0.5;0.77;0.95;1$.}
\label{fz}
\end{figure*}

The Majumdar-Papapetrou multiblack hole solution is \cite{majumdar1947,papapetrou1947}
\begin{eqnarray}
 ds^2=-\Omega^{-2}dt^2+\Omega^2(dx^2+dy^2+dz^2),
 \label{metricMP}
\end{eqnarray}
where $\Omega=1+\sum_{i}\frac{m_i}{r_i}$ and
$$r_i=\sqrt{(x-x_i)^2+(y-y_i)^2+(z-z_i)^2}.$$
"i" is number of black holes, $m_i$ is the mass of the i-th black hole, $x_i$, $y_i$, $z_i$ is located of the i-th black hole. Here we will consider the case when $i=2$ and use a cylindrical coordinate system in which black holes are located on the z axis \cite{chernov2023}
\begin{eqnarray}
 x=r\cos\phi,\quad
 y=r\sin\phi,\quad
 z=z.
\end{eqnarray}
The plane $z=0$ is the equatorial plane. Then Majumdar-Papapetrou solution is rewritten in the form \cite{chernov2023,wunsch2013}
\begin{eqnarray}
 ds^2=-\Omega^{-2}dt^2+\Omega^2(dr^2+r^2d\phi^2+dz^2),
 \label{metricMPcyl}
\end{eqnarray}
where $r_i=\sqrt{r^2+(z-z_i)^2}$ and $x_i=y_i=0$. In the appendix you can find the metric coefficients and the symbols of Christoffel for this metric (\ref{metricMPcyl}). The function $\Omega$ depends on two variables, $r$, $z$. We will need derivatives of this function, $\Omega$, that are equal to
\begin{eqnarray}
 \frac{\partial\Omega}{\partial z}=-\sum_i\frac{m_i(z-z_i)}{(r^2+(z-z_i)^2)^{3/2}},\\
 \frac{\partial\Omega}{\partial r}=-\sum_i\frac{m_ir}{(r^2+(z-z_i)^2)^{3/2}}.
\end{eqnarray}
Equation of motion for photon is described by equation \cite{wunsch2013}
\begin{eqnarray}
 \dot{r}^2+\dot{z}^2+V=E^2
 \label{EqMotion}
\end{eqnarray}
with the effective potential
\begin{eqnarray}
 V=\frac{L^2}{r^2\Omega^4}
\end{eqnarray}
and two constants of motion
\begin{eqnarray}
 E=-p_t,\quad L=p_\phi,
\end{eqnarray}
where $E$ is energy and $L$ is angular momentum of photon. Here we will consider the case in which $m_1\neq m_2$. The case when $m_1=m_2$ was considered in paper \cite{wunsch2013}.
Without loss of generality, consider the case when $z_1=-z_2=l$. The distance between black holes equals 2l.

As follows from equation (\ref{EqMotion}), circular orbits will satisfy the equations,
$\frac{\partial V}{\partial r}=\frac{\partial V}{\partial z}=0$ that reduce to the form
\begin{eqnarray}
 \frac{m_1}{m_2}=\frac{l+z}{l-z}(1-\frac{4zl}{r^2+(z+l)^2})^{3/2},\nonumber\\
 1+\frac{m_1((z-l)^2-r^2)}{(r^2+(z-l)^2)^{3/2}}+
 \frac{m_2((z+l)^2-r^2)}{(r^2+(z+l)^2)^{3/2}}=0.
 \label{dVdzdr}
\end{eqnarray}
The first equation of (\ref{dVdzdr}) shows that circular orbits exist only in the range $-l<z<l$. Circular orbits in the equatorial plane can only exist for $m_1=m_2$ (see \cite{wunsch2013}). We will rewrite the first equation of (\ref{dVdzdr}) in the form
\begin{eqnarray}
r^2=\frac{4zl}{1-(\frac{m_1}{m_2}\frac{l-z}{l+z})^{2/3}}-(z+l)^2.
\label{r2z}
\end{eqnarray}
This equation describes the motion of photons for $z=const$. Substituting the equation (\ref{r2z}) into the second equation (\ref{dVdzdr}) results in
\begin{eqnarray}
 f(z)=(4lA-(z+l)^2)\left(\frac{m_1}{(A-z)^{3/2}}+\frac{m_2}{A^{3/2}}\right)-\nonumber\\
 -2l^{3/2}\left(2+\frac{m_1}{\sqrt{lA-lz}}+\frac{m_2}{\sqrt{lA}}\right),
\end{eqnarray}
where
\begin{eqnarray}
 A=\frac{z}{1-\left(\frac{m_1}{m_2}\right)^{2/3}\left(\frac{l-z}{l+z}\right)^{2/3}}.
\end{eqnarray}
Without loss of generality, we will consider the case when, $m_1<m_2$ and let's introduce the notation, $q=\frac{m_1}{m_2}<1$. The opposite case is treated similarly.

In the figure (\ref{fz}) you can find the dependence of the function f on z for the following parameters: $q=1$ - blue solid line, $q=0.95$ - green dashed line, $q=0.77$ - red dash-dotted line, $q=0.5$ - black dotted line. The height of
circular orbits is determined by the roots of the equation, $f(z)=0$. For black holes with equal masses, $m_1=m_2$, there may be zero, one or two circular orbits \cite{wunsch2013}.
For black holes of unequal masses, $m_1\neq m_2$, there may be two or four circular orbits (see fig. (\ref{fz})). The radii of these orbits are determined by the equation, (\ref{r2z}).

\section{Shadows and photon rings of binary black holes}

\begin{figure*}
\includegraphics[scale=0.5]{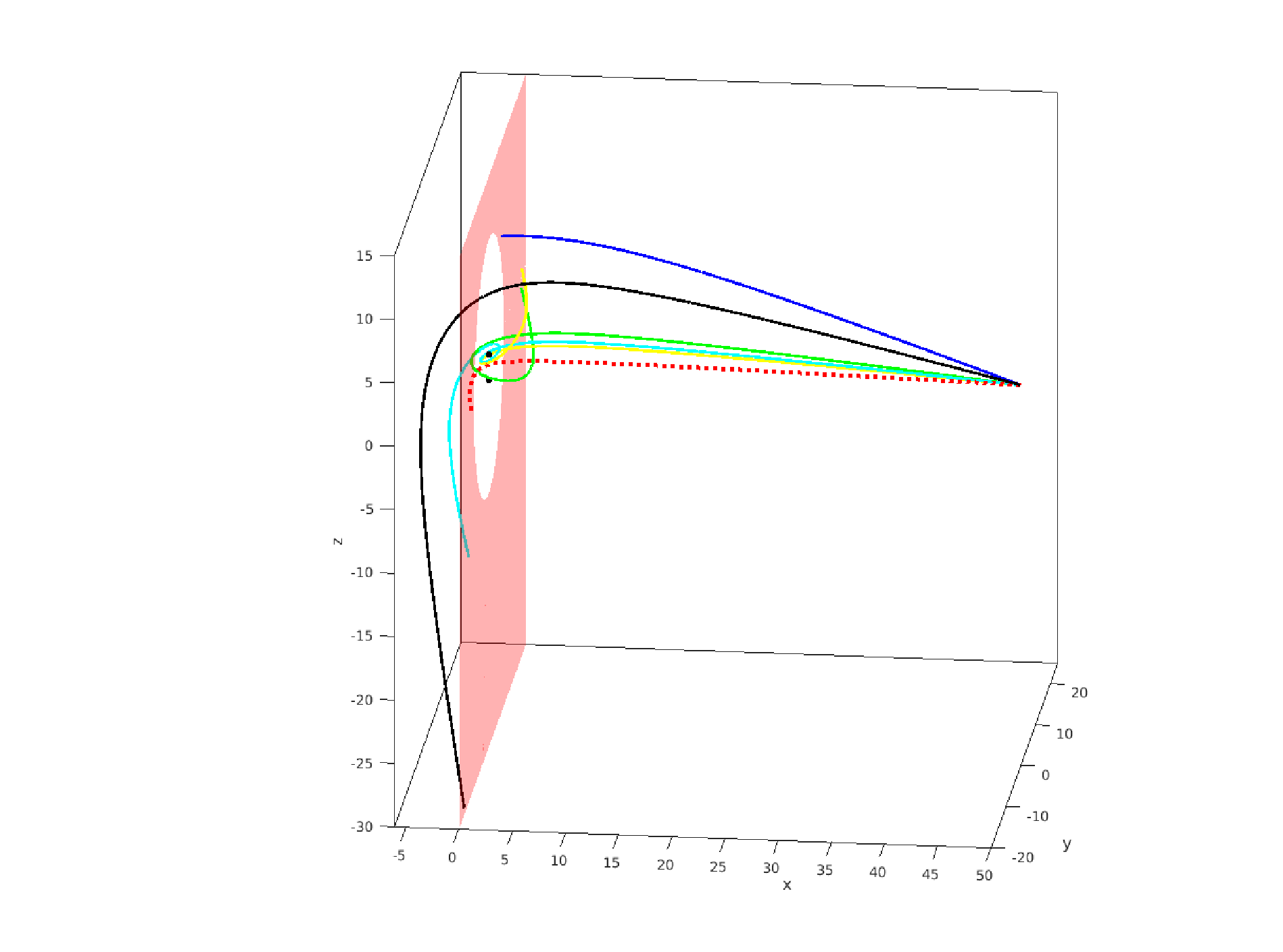}
\includegraphics[scale=0.5]{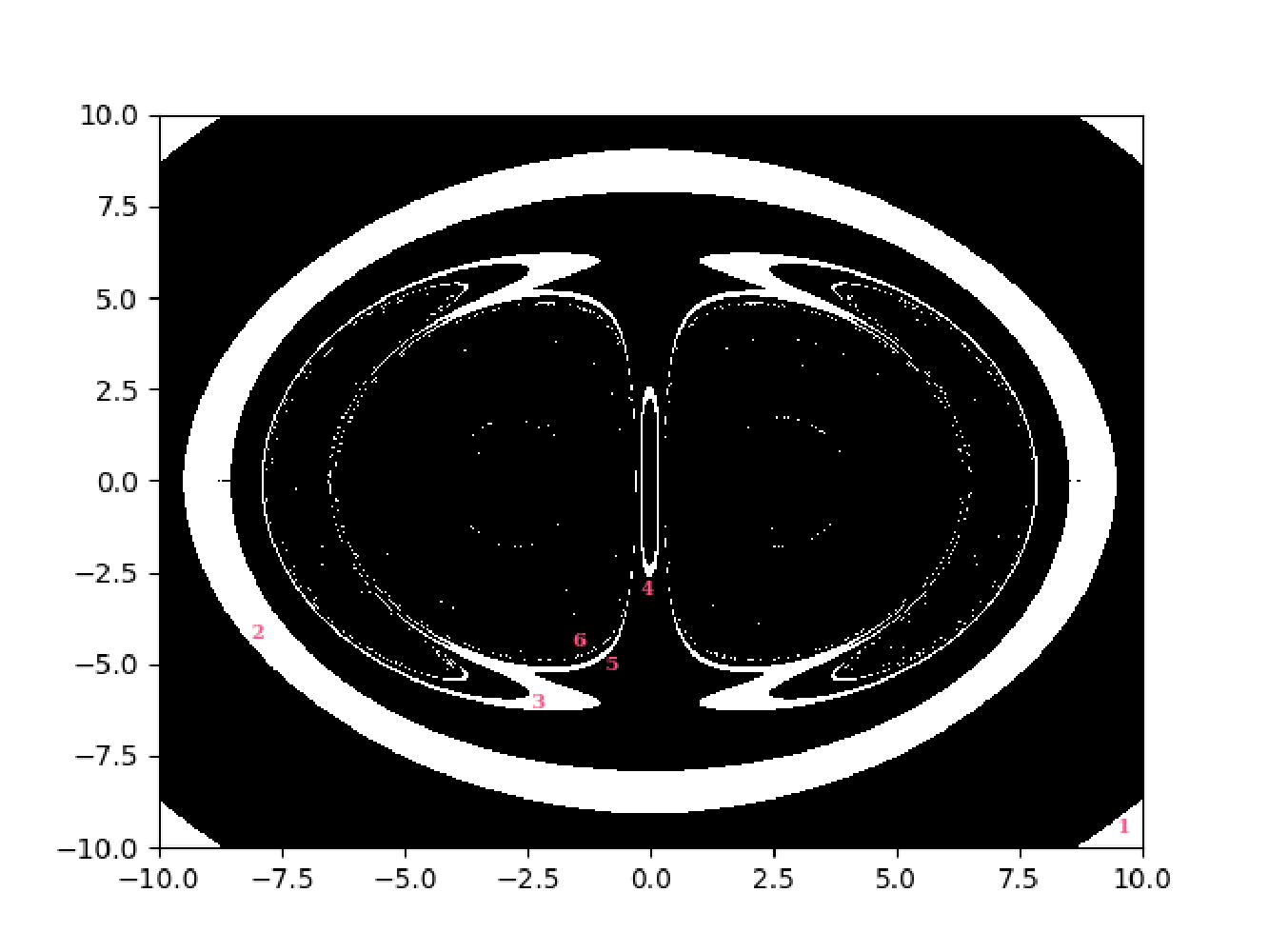}
\caption{In the figure you can see the image of a binary black holes and the trajectories of photons that correspond to each photon ring.}
\label{im4}
\end{figure*}

In this section, we will present the results of modeling the shadows and photon rings of binary black holes for a geometrically thin disk. It's assume that the thin disk is located in the plane, $r=0$ and $\phi=0,\pi$ (see fig. (\ref{im4})) and emits isotropically in all direction . The inner radius of the thin disk was set equal to $r_{in}=10$ for the observer's inclination $i=90^\circ$ and $r_{in}=20$ for another observer's inclination and it was assumed that the disk extends to infinity.

\begin{figure*}
\includegraphics[scale=0.22]{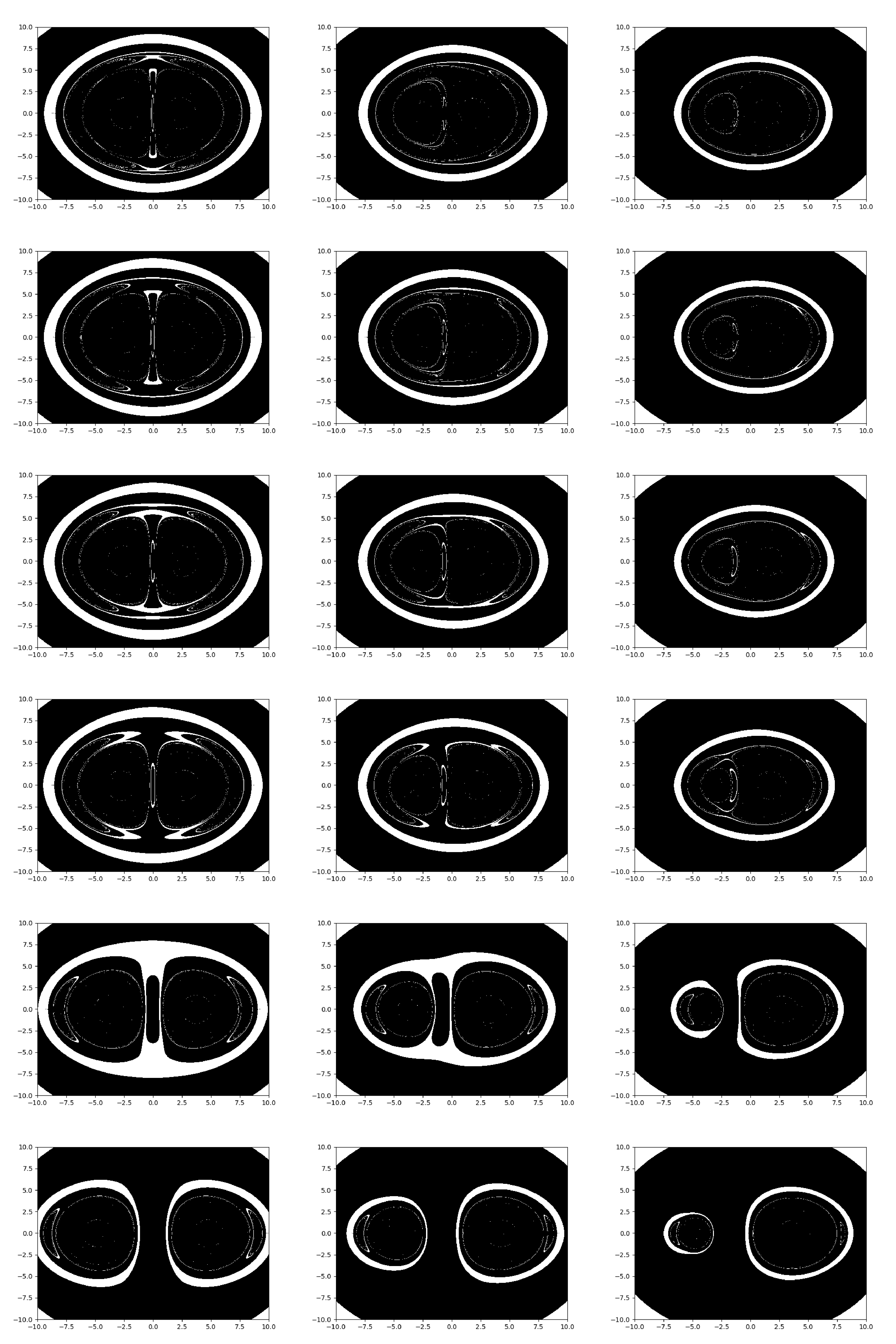}
\caption{The figure shows images of binary black holes in the case when the observer is located at an angle of $90^\circ$ degrees. The mass of black hole from left to right is equal $m_1=1$, $m_1=0.7$ and $m_1=0.4$. The mass of the second black hole is equal to $m_2=1$. The distance between black holes from top to bottom is equal $l=0.7;0.8;0.9;1;2;3$.}
\label{im1}
\end{figure*}

\begin{figure*}
\includegraphics[scale=0.22]{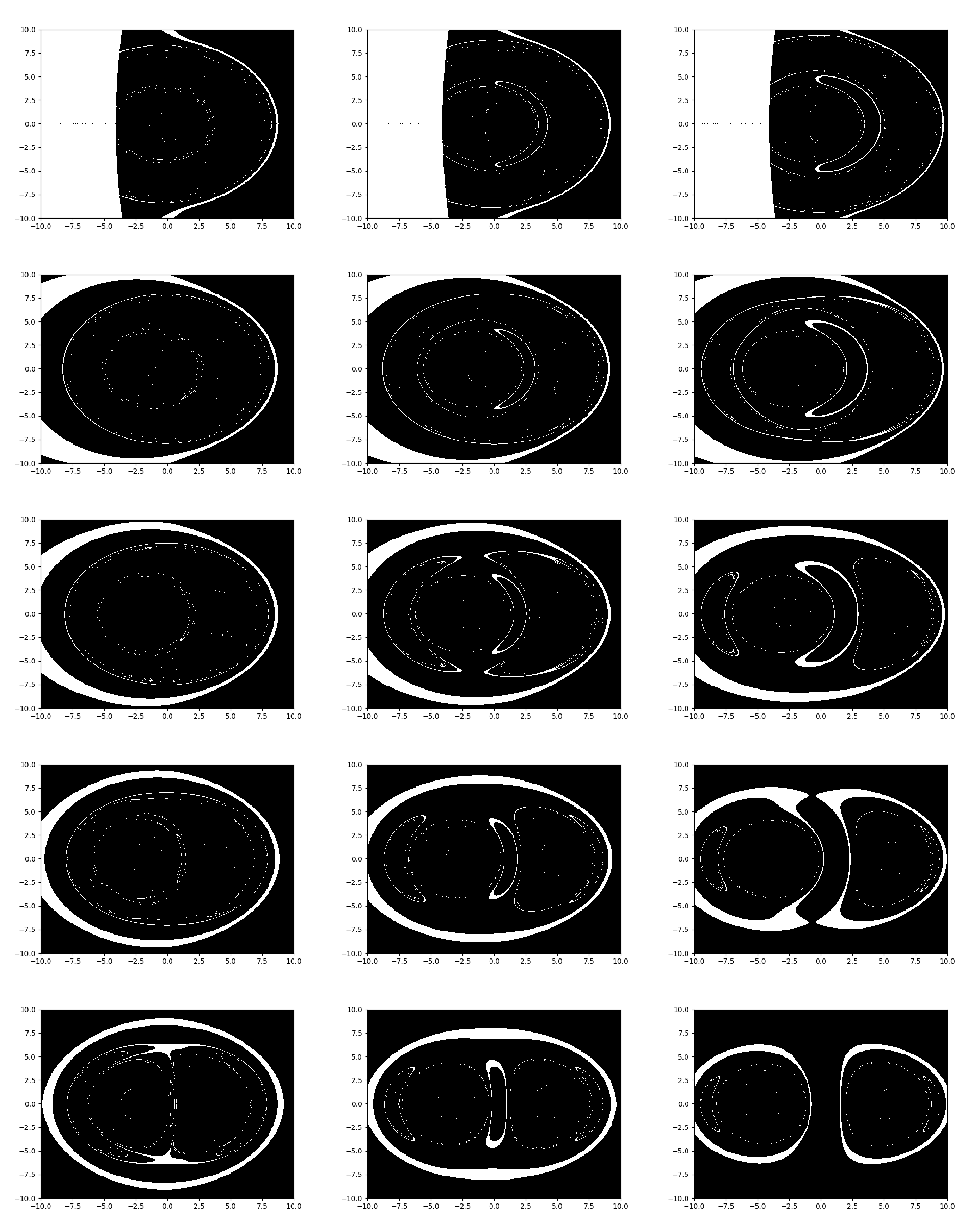}
\caption{The figure shows images of binary black holes in the case when the masses of the black hole are equal $m_1=m_2=1$. The left column corresponds to the case when the distance between black holes is equal $l=1$, the middle column is $l=2$, the right column is $l=3$. The angle of inclination of the observer to black holes from top to bottom is equal to $i=10^\circ;30^\circ;45^\circ;60^\circ;80^\circ$.}
\label{im2}
\end{figure*}

\begin{figure*}
\includegraphics[scale=0.22]{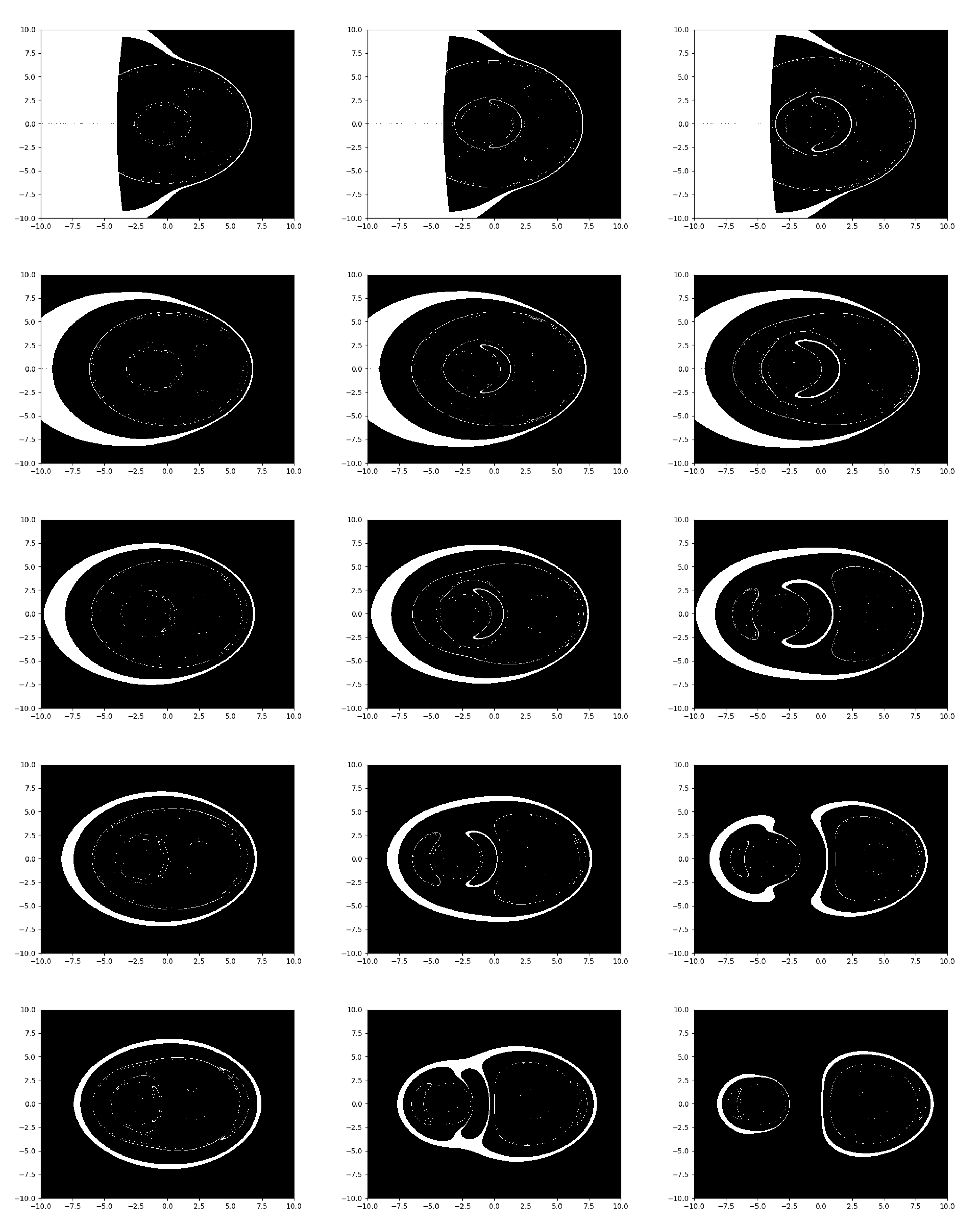}
\caption{The figure shows images of binary black holes in the case when the masses of the black hole are equal $m_1=0.5$ and $m_2=1$. The left column corresponds to the case when the distance between black holes is equal $l=1$, the middle column is $l=2$, the right column is $l=3$. The angle of inclination of the observer to black holes from top to bottom is equal to $i=10^\circ;30^\circ;45^\circ;60^\circ;80^\circ$.}
\label{im3}
\end{figure*}

Let's consider an observer who is located far enough away from binary black holes. Then we define the celestial coordinates of the observer as
\begin{eqnarray}
\alpha=-\frac{\sqrt{r^2+z^2}p^{(\phi)}}{p^{(t)}}\\
\beta=-\frac{\sqrt{r^2+z^2}}{p^{(t)}}(p^{(z)}\sin i-p^{(r)}\cos i),
\end{eqnarray}
where $i$ is the angle between the observer and the line connecting the binary black holes. When the observer is located at the equatorial plane ($z=0$) then the angle $i$ equals, $i=90^\circ$. The index in parentheses means the tetrad's index. The tetrad vectors are given in the appendix. In the numerical calculations, the observer was located at a distance $r_o=50$ and at inclination, $i=10^\circ,30^\circ,45^\circ,60^\circ,80^\circ,90^\circ$.

Photons are emitted by a thin disk and move along null geodesics. The null geodesics for photon trajectories are described by a system of eight coupled first-order equations
\begin{eqnarray}
 \frac{dx^\alpha}{d\lambda}=p^\alpha,\\
 \frac{dp^\alpha}{d\lambda}=-\Gamma^\alpha_{\mu\nu}p^\mu p^\nu.
\end{eqnarray}
The null geodesic equations were numerically
solved using the fourth-order Runge–Kutta method. The shadows and photon rings were constructed using the ray-tracing method. The images are shown on a scale from –10 to +10 in units of $GM/c^2$ with a resolution of $2500\times2500$ pixels.

In the figures (\ref{im1}-\ref{im3}) you can see images of binary black holes depending on the parameters of the problem. The images consist of dark areas - shadows and bright closed lines - photon rings. In Figure (\ref{im1}), the observer is located at an angle of $i=90^\circ$ degrees relative to black holes. The left column corresponds to the case when the mass of the black hole is equal to, $m_1=1$, the middle column - $m_1=0.7$, the right column - $m_1=0.4$. The mass of second black hole is equal to, $m_2=1$. The figures from top to bottom correspond to the case when the distance between black holes is equal to, $l=0.7;0.8;0.9;1;2;3$.

The figures (\ref{im2}) and (\ref{im3}) correspond to the case when the observer is located at an angle $10^\circ,30^\circ,45^\circ,60^\circ,80^\circ$ (from top to bottom) to a binary black hole. The left column corresponds to the case when the distance between of binary black holes is equal to, $l=1$, the middle column - $l=2$, the left column - $l=3$. The figure (\ref{im2}) corresponds to the case when black holes are of equal masses, $m_1=m_2=1$, and the figure (\ref{im2}) corresponds to the case when the mass of one black hole is twice as large as the other, $m_1=0.5$, $m_2=1$.

To understand where the photon rings are located in the image and what order they are,
it is necessary to depict the trajectories of photons around a binary black hole. To do this, 
let's take for example the image of a binary black hole with the following parameters,
$m_1=m_2=l=1$ and $i=90^\circ$. This image is shown in the figure (\ref{im4}). When creating this image using backward ray tracing, photons with parameters, $-10<\alpha<10$ and $-10<\beta<10$ were launched. Photons that have made a half-turn or more around one or two black holes at once belong to photon rings. Examples of such trajectories are shown in the figure (\ref{im4}). In total, six photon trajectories are depicted.

The blue solid trajectory that leaves from point 1 ($\alpha=-10$, $\beta=10$) in the figure (\ref{im4}) 
does not make a half-turn (or revolution) around black holes and corresponds to the image of the disk in the figure (\ref{im4}). The black solid trajectory that leaves from point 2 ($\alpha=-4.072$, $\beta=-8$) in the figure (\ref{im4}) makes a half-turn around two black holes and corresponds to the first photon ring.
The green solid trajectory that leaves from point 3 ($\alpha=-6$, $\beta=-2$) in the figure (\ref{im4}) 
makes a revolution around two black holes corresponds to the second photon ring. Thus, the eyebrowlike structure corresponds to photon rings that are nested sequentially into each other. These photon rings correspond to the trajectories of a photon around two black holes at once as one. In other words, those photons that revolve around two black holes as a whole form a eyebrowlike structure.

A dotted red trajectory that leaves point 4 ($\alpha=-2.552$, $\beta=0$) passed between two black holes and makes one half-turn around one of the black holes. There is no analogue of such a photon trajectory around a single black hole. These photons form a photon ring between black holes.

A solid yellow trajectory that leaves point 5 ($\alpha=-4.784$, $\beta=-1$) makes a revolution around one black hole, forming photon rings. A solid blue trajectory that comes out of point 6
($\alpha=-4.784$, $\beta=-1.464$) makes one and a half turns also around one black hole, forming the
following photon rings. These photon trajectories form a system of photon rings nested inside each
other and similar to the system of photon rings around a single black hole.

\section{Conclusions}

In this paper, we calculated the null trajectories of photons and constructed images of binary black holes depending on the parameters of the problem: the mass of black holes, the distance between them,
the angle of inclination of the observer to the black holes. These images contain both shadows and photon rings.

We admit that this solution may be far from a realistic situation, nevertheless, these studies will help us to better understand the physical processes of the physics of binary black holes.

\section{Appendix}

Metric coefficients are
\begin{eqnarray}
 g_{tt}=-\frac{1}{\Omega^2},\quad g_{rr}=\Omega^2,\quad
 g_{zz}=\Omega^2,\quad g_{\phi\phi}=r^2\Omega^2\nonumber\\
 g^{tt}=-\Omega^2,\quad g^{rr}=\frac{1}{\Omega^2},\quad
 g^{zz}=\frac{1}{\Omega^2},\quad g^{\phi\phi}=\frac{1}{r^2\Omega^2}.\nonumber
\end{eqnarray}
The nonzero Christoffel symbols are
\begin{eqnarray}
 \Gamma^t_{tr}=-\frac{1}{2\Omega^2}\frac{\partial\Omega^2}{\partial r},\quad \Gamma^t_{tz}=-\frac{1}{2\Omega^2}\frac{\partial\Omega^2}{\partial z},\nonumber\\
 \Gamma^r_{tt}=-\frac{1}{2\Omega^6}\frac{\partial\Omega^2}{\partial r},\quad \Gamma^r_{rr}=\frac{1}{2\Omega^2}\frac{\partial\Omega^2}{\partial r},\nonumber\\
 \Gamma^r_{\phi\phi}=-\frac{1}{2\Omega^2}\frac{\partial (r^2\Omega^2)}{\partial r},\quad \Gamma^r_{zz}=-\frac{1}{2\Omega^2}\frac{\partial\Omega^2}{\partial r},\nonumber\\
 \Gamma^\phi_{r\phi}=\frac{1}{2r^2\Omega^2}\frac{\partial (r^2\Omega^2)}{\partial r},\quad \Gamma^\phi_{\phi z}=\frac{1}{2r^2\Omega^2}\frac{\partial (r^2\Omega^2)}{\partial z},\nonumber\\
 \Gamma^z_{tt}=-\frac{1}{2\Omega^6}\frac{\partial\Omega^2}{\partial z}\quad\Gamma^r_{rz}=\frac{1}{2\Omega^2}\frac{\partial\Omega^2}{\partial z},\nonumber\\
 \Gamma^z_{rr}=-\frac{1}{2\Omega^2}\frac{\partial\Omega^2}{\partial z},\quad \Gamma^z_{rz}=\frac{1}{2\Omega^2}\frac{\partial\Omega^2}{\partial r},\nonumber\\ \Gamma^z_{\phi\phi}=-\frac{r^2}{2\Omega^2}\frac{\partial\Omega^2}{\partial z}\quad
 \Gamma^z_{zz}=\frac{1}{2\Omega^2}\frac{\partial\Omega^2}{\partial z}.\nonumber
\end{eqnarray}
The local tetrad is defined as follows ($\alpha=t,r,\phi,z$)
\begin{eqnarray}
 e^{(t)}_{\alpha}=\left(\frac{1}{\Omega},0,0,0\right),\quad e^{(r)}_{\alpha}=\left(0,\Omega,0,0\right),\nonumber\\
 e^{(\phi)}_{\alpha}=\left(0,0,r\Omega,0\right),\quad e^{(z)}_{\alpha}=\left(0,0,0,\Omega\right).\nonumber
\end{eqnarray}
The transition from one coordinate system to another is performed according to the formulas
$$p^{(a)}=e^{(a)}_\alpha p^\alpha,\quad p^\alpha=e^\alpha_{(a)}p^{(a)}.$$


\begin{thebibliography}{99}

\bibitem{abbott2016} B. P. Abbott et al., Phys. Rev. Lett. {\bf 116}, 241102 (2016).

\bibitem{malinovsky2023} A. M. Malinovsky, E. V. Mikheeva, Astronomy Reports, {\bf 67}, 685 (2023).

\bibitem{titarchuk2023} L. Titarchuk, E. Seifina, and C. Shrader, Astron. Astrophys. {\bf 671}, A159 (2023).

\bibitem{eht2019} The Event Horizon Telescope Collaboration et. al. ApJL {\bf 875}, L1 (2019).

\bibitem{eht2022} The Event Horizon Telescope Collaboration et. al. ApJL, {\bf 930}, L12 (2022)

\bibitem{kardashev2014} N. S. Kardashev, I. D. Novikov, V. N. Lukash, S. V. Pilipenko, E. V. Mikheeva, D. V. Bisikalo, D. S. Wiebe, A. G. Doroshkevich, A. V. Zasov, I. I. Zinchenko, P. B. Ivanov, V. I. Kostenko, T. I. Larchenkova, S. F. Likhachev, I. F. Malov, V. M. Malofeev, A. S. Pozanenko, A. V. Smirnov, A. M. Sobolev, A. M. Cherepashchuk, Yu. A. Shchekinov, Physics-Uspekhi {\bf 57}, 1199 (2014).

\bibitem{cunha2018} P.V.P. Cunha, C.A.R. Herdeiro and M.J. Rodriguez, Phys. Rev. D {\bf 98}, 044053 (2018).

\bibitem{yumoto2012} A. Yumoto, D. Nitta, T. Chiba, N. Sugiyama, Phys. Rev. D {\bf 86}, 103001 (2012).

\bibitem{shipley2016} J.O. Shipley, S.R. Dolan, Class. Quantum. Grav. {\bf 33}, 175001 (2016).

\bibitem{nitta2011} D. Nitta, T. Chiba, N. Sugiyama, Phys. Rev. D {\bf 84}, 063008 (2011).

\bibitem{majumdar1947} S. D. Majumdar, Phys. Rev. {\bf 72}, 390 (1947).

\bibitem{papapetrou1947} A. Papapetrou, Proc. Roy. Irish Acad. {\bf 51}, 191 (1947).

\bibitem{chernov2023} S.V. Chernov, Gravitation and Cosmology, {\bf 29}, 437 (2023); arXiv:2306.03826

\bibitem{wunsch2013} A. Wunsch, T. Muller, D. Weiskopf, G. Wunner, Phys. Rev. D {\bf 87}, 024007 (2013).

\end{thebibliography}
\end{document}